\documentclass[nofootinbib,prd,twocolumn,showpacs,showkeys,preprintnumbers]{revtex4-1}
\usepackage{hyperref,amssymb,amsmath,mathrsfs,bm,graphicx}
\begin{document}
\title {Expansion--free dissipative fluid spheres: Analytical models}
\author{L. Herrera}
\email{lherrera@usal.es}
\affiliation{Instituto Universitario de F\'isica
Fundamental y Matem\'aticas, Universidad de Salamanca, Salamanca 37007, Spain.}
\author{A. Di Prisco}
\email{adiprisco56@gmail.com}
\affiliation{Escuela de F\'{\i}sica, Facultad de Ciencias,
Universidad Central de Venezuela, Caracas 1050, Venezuela.}
\author{J. Ospino}\affiliation{Departamento de Matem\'atica Aplicada and Instituto Universitario de F\'isica
Fundamental y Matem\'aticas, Universidad de Salamanca, Salamanca 37007, Spain.}
\email{j.ospino@usal.es}
\date{\today}
\begin{abstract}
 We search  exact analytical solutions of  spherically symmetric dissipative fluid distributions satisfying the vanishing expansion condition (vanishing expansion scalar $\Theta$). To do so we shall impose additional restrictions allowing the integration of the field equations. A detailed analysis  of the obtained solutions, their prospective applications  to astrophysical scenarios, as well as alternative approaches to obtain new solutions, are discussed.

\end{abstract}
\date{\today}
\maketitle

\section{Introduction}
Many years ago V. Skripkin published a paper describing the evolution of a spherically symmetric distribution of incompressible non--dissipative fluid, following a central explosion \cite{Skripkin} (see also  \cite{Mac}).  Although not explicitly stated in the Skripkin paper, his model implies the vanishing of the expansion scalar. The interest of this kind of  models stems from the fact  that  the expansion-free condition necessarily implies the appearance of a cavity around the center, suggesting   that they  might be relevant for the modeling, among other phenomena,  of voids observed at cosmological scales.

A general study on shearing expansion-free  spherical fluid evolution (including pressure anisotropy) was carried out in  \cite{H1}, where the unavoidable appearance of a cavity  surrounding the center in expansion--free solutions was explained as consequence of the fact  that the $\Theta=0$ condition requires that the innermost  shell of fluid should be away from the centre,  initiating therefrom  the formation of a cavity.

Later on in \cite{H2} it was shown that the Skripkin model is ruled out if Darmois junctions conditions are imposed (both boundary surfaces are thin shells for this model).

Thus,  the evolution  will proceed expansion--free if  the decrease (increase) of the areal radius of  the outer boundary surface of the fluid distribution is compensated by a decrease (increase) of the areal radius of the boundary of the cavity, (see \cite{H1} for a detailed discussion on this issue).

General results  regarding expansion--free fluids may be found in \cite{19, 20}, and extensions to other  theories of gravitation  and/or charged  fluids and/or different kind of symmetries have been discussed in \cite{4, 15, 3, 5, 6b,7b} and references therein.

The problem of (in)stability under  the expansion--free condition  was addressed by the first time  in \cite{H3}, and afterwards this problem has been addressed within the context of modified theories of gravity in \cite{7, 8, 9,10,12,18} and references therein.

Besides the solutions presented in \cite{H1}, exact solutions describing expansion--free fluids may be found in \cite{11a,11,13,14,16,17,21}.

Due to the interest raised by expansion--free fluids, it is our purpose in this work to find exact  analytical solutions to Einstein equations  for dissipative expansion--free fluids.
These solutions will be found by imposing additional restrictions allowing to the full integration of field equations. Some of these restrictions are endowed with a distinct physical meaning, while others are just of heuristic nature. Among the former, stand out the vanishing complexity factor condition  and the quasi--homologous evolution.  A discussion on all the presented models is  brought out  in the last section.

\section{Relevant physical  and geometric variables, field equations and  junction conditions}
We consider  spherically symmetric distributions  of
fluid, bounded from the exterior by a spherical surface $\Sigma^{(e)}$. The fluid is
assumed to be locally anisotropic (principal stresses unequal) and undergoing dissipation in the
form of heat flow (to model dissipation in the diffusion approximation).

The motivation to include dissipation is provided by the well known fact that  dissipation due to the emission of massless particles (photons and/or neutrinos) seems  to be  the only plausible mechanism to carry away the bulk of the binding energy of the collapsing star, leading to a neutron star or a black hole. On the other hand, the diffusion approximation is in general very sensible, since the mean free path of particles responsible for the propagation of energy in stellar interiors  is in general very small as compared with the typical length of the object. Furthermore, the observational data collected from supernova 1987A indicates that the regime of radiation transport prevailing during the emission process is closer to the diffusion approximation than to the streaming out limit. Thus, even though in many other circumstances, the mean free path of particles transporting energy may be large enough as to justify the  free streaming approximation, there are many physically meaningful scenarios where the diffusion approximation is justified.

The fluid under consideration is anisotropic in the pressure, this is well justified since local isotropy, as it has been shown in recent years, is a too stringent condition which may excessively constrain modeling of self-gravitating objects. Furthermore, local anisotropy of pressure may be caused by a large variety of physical phenomena, of the kind we expect in compact objects \cite{rep}. More so, as it has been recently shown \cite{ps},  physical processes of the kind expected in stellar evolution will always tend to produce pressure anisotropy, even if the system is initially assumed to be isotropic.  Since any equilibrium configuration is the final stage of a dynamic  regime,  there is no reason to think that the acquired anisotropy during this dynamic process   would disappear in the final equilibrium state, and therefore the resulting configuration, even if  initially  had isotropic pressure, should in principle exhibit pressure anisotropy.

As we mentioned in the introduction, the expansion-free models  present an internal vacuum cavity. We shall denote by  $\Sigma^{(i)} $ the boundary surface between the cavity (inside which we have a Minkoswki spacetime) and the fluid.

Choosing comoving coordinates inside $\Sigma^{(e)}$, the general
interior metric can be written
\begin{equation}
ds^2=-A^2dt^2+B^2dr^2+R^2(d\theta^2+\sin^2\theta d\phi^2),
\label{1}
\end{equation}
where $A$, $B$ and $R$ are functions of $t$ and $r$ and are assumed
positive. We number the coordinates $x^0=t$, $x^1=r$, $x^2=\theta$
and $x^3=\phi$. Observe that $A$ and $B$ are dimensionless, whereas $R$ has the same dimension as $r$.

The  energy momentum tensor  in the canonical form, reads
\begin{equation}
T_{\alpha \beta} = {\mu} V_\alpha V_\beta + P h_{\alpha \beta} + \Pi_{\alpha \beta} +
q _\beta V_\alpha + q_\alpha V_\beta, \label{Tab}
\end{equation}
with
$$ P=\frac{P_{r}+2P_{\bot}}{3}, \qquad h_{\alpha \beta}=g_{\alpha \beta}+V_\alpha V_\beta,$$

$$\Pi_{\alpha \beta}=\Pi\left(K_\alpha K_\beta - \frac{1}{3} h_{\alpha \beta}\right), \qquad \Pi=P_{r}-P_{\bot},$$
where $\mu$ is the energy density, $P_r$ the radial pressure,
$P_{\perp}$ the tangential pressure, $q^{\alpha}$ the heat flux, $V^{\alpha}$ the four--velocity of the fluid,
and $K^{\alpha}$ a unit four--vector along the radial direction.
Since we are considering comoving observers, we have
\begin{eqnarray}
V^{\alpha}&=&A^{-1}\delta_0^{\alpha}, \;\;
q^{\alpha}=qK^{\alpha}, \;\;
K^{\alpha}=B^{-1}\delta^{\alpha}_1.
\end{eqnarray}

 These quantities
satisfy
\begin{eqnarray}
V^{\alpha}V_{\alpha}=-1, \;\; V^{\alpha}q_{\alpha}=0, \;\; K^{\alpha}K_{\alpha}=1,\;\;
K^{\alpha}V_{\alpha}=0.
\end{eqnarray}

It is worth noticing that we do not explicitly add bulk or shear viscosity to the system because they
can be trivially absorbed into the radial and tangential pressures, $P_r$ and
$P_{\perp}$  (in $\Pi$). Also we do not explicitly  introduce  dissipation in the free streaming approximation since it can be absorbed in $\mu, P_r$ and $q$.

The acceleration $a_{\alpha}$, the expansion $\Theta$, and the   shear $\sigma_{\alpha\beta}$  of the fluid are
given by
\begin{equation}
a_{\alpha}=V_{\alpha ;\beta}V^{\beta}, \;\;
\Theta={V^{\alpha}}_{;\alpha}, \label{4b}
\end{equation}

\begin{equation}
\sigma_{\alpha\beta}=V_{(\alpha
;\beta)}+a_{(\alpha}V_{\beta)}-\frac{1}{3}\Theta h_{\alpha\beta}.
\label{4a}
\end{equation}

From  (\ref{4b}) we have for the  four--acceleration and its scalar $a$,
\begin{equation}
a_\alpha=a K_\alpha, \;\; a=\frac{A^{\prime}}{AB}, \label{5c}
\end{equation}
and for the expansion
\begin{equation}
\Theta=\frac{1}{A}\left(\frac{\dot{B}}{B}+2\frac{\dot{R}}{R}\right),
\label{5c1}
\end{equation}
where the  prime stands for $r$
differentiation and the dot stands for differentiation with respect to $t$.

We obtain
for the shear (\ref{4a}) its non zero components
\begin{equation}
\sigma_{11}=\frac{2}{3}B^2\sigma, \;\;
\sigma_{22}=\frac{\sigma_{33}}{\sin^2\theta}=-\frac{1}{3}R^2\sigma,
 \label{5a}
\end{equation}
and its scalar
\begin{equation}
\sigma^{\alpha\beta}\sigma_{\alpha\beta}=\frac{2}{3}\sigma^2,
\label{5b}
\end{equation}
where
\begin{equation}
\sigma=\frac{1}{A}\left(\frac{\dot{B}}{B}-\frac{\dot{R}}{R}\right).\label{5b1}
\end{equation}

Einstein's field equations for the interior spacetime (\ref{1}) \begin{equation}
G_{\alpha\beta}=8\pi T_{\alpha\beta},
\label{2}
\end{equation}
are given in the Appendix A.

Thus in the most general case (locally anisotropic and dissipative) we have available four field equations (\ref{12}--\ref{15}) for  seven variables, namely  $A$, $B$, $R$, $\mu$, $P_r$, $P_{_\perp}$, and  $q$. Since we are going to consider expansion--free systems we have the additional condition $\Theta=0$. Evidently, in order to find specific models (to close the system of equations) we need to provide additional information, which could be given in the form of constitutive equations for $q$, and/or equations of state for both pressures, and/or any other type of constraint on physical and/or metric variables.  In this work  we shall impose (among others) the vanishing complexity factor condition and the quasi--homologous evolution condition.

Next, the mass function $m(t,r)$ introduced by Misner and Sharp
\cite{Misner} (see also \cite{Cahill}) reads
\begin{equation}
m=\frac{R^3}{2}{R_{23}}^{23}
=\frac{R}{2}\left[\left(\frac{\dot R}{A}\right)^2-\left(\frac{R^{\prime}}{B}\right)^2+1\right].
 \label{17masa}
\end{equation}

To study the dynamical properties of the system it is convenient to  introduce
the proper time derivative $D_T$
given by
\begin{equation}
D_T=\frac{1}{A}\frac{\partial}{\partial t}, \label{16}
\end{equation}
and the proper radial derivative $D_R$,
\begin{equation}
D_R=\frac{1}{R^{\prime}}\frac{\partial}{\partial r}, \label{23a}
\end{equation}
where $R$ defines the areal radius of a spherical surface inside $\Sigma^{(e)}$ (as
measured from its area).

Using (\ref{16}) we can define the velocity $U$ of the collapsing
fluid as the variation of the areal radius with respect to proper time, i.e.
\begin{equation}
U=D_TR. \label{19}
\end{equation}
Then (\ref{17masa}) can be rewritten as
\begin{equation}
E \equiv \frac{R^{\prime}}{B}=\left(1+U^2-\frac{2m}{R}\right)^{1/2}.
\label{20x}
\end{equation}
With (\ref{23a}) we can express (\ref{17a}) as
\begin{equation}
4\pi q=E\left[\frac{1}{3}D_R(\Theta-\sigma)
-\frac{\sigma}{R}\right].\label{21a}
\end{equation}

Using (\ref{12})-(\ref{14}) with (\ref{16}) and (\ref{23a}) we obtain from
(\ref{17masa})
\begin{eqnarray}
D_Tm=-4\pi\left(P_r U+qE\right)R^2,
\label{22Dt}
\end{eqnarray}
and
\begin{eqnarray}
D_Rm=4\pi\left(\mu+q\frac{U}{E}\right)R^2,
\label{27Dr}
\end{eqnarray}
which implies
\begin{equation}
m=4\pi\int^{r}_{0}\left(\mu +q\frac{U}{E}\right)R^2R^\prime d\tilde r. \label{27intcopy}
\end{equation}
(assuming a regular centre to the distribution, so $m(0)=0$).

\subsection{The exterior spacetime and junction conditions}
Outside $\Sigma^{(e)}$ we assume we have the Vaidya
spacetime (i.e.\ we assume all outgoing radiation is massless),
described by
\begin{equation}
ds^2=-\left[1-\frac{2M(v)}{{ \bf r}}\right]dv^2-2d{\bf r}dv+{\bf r}^2(d\theta^2
+\sin^2\theta
d\phi^2) \label{1int},
\end{equation}
where $M(v)$  denotes the total mass,
and  $v$ is the retarded time.

The matching of the full nonadiabatic sphere  (including viscosity) to
the Vaidya spacetime, on the surface $r=r_{\Sigma^{(e)}}=$ constant was discussed in
\cite{chan1}. From the continuity of the first and second differential forms it follows (see \cite{chan1} for details)
\begin{equation}
m(t,r)\stackrel{\Sigma^{(e)}}{=}M(v), \label{junction1}
\end{equation}
and
\begin{widetext}
\begin{eqnarray}
2\left(\frac{{\dot R}^{\prime}}{R}-\frac{\dot B}{B}\frac{R^{\prime}}{R}-\frac{\dot R}{R}\frac{A^{\prime}}{A}\right)
\stackrel{\Sigma^{(e)}}{=}-\frac{B}{A}\left[2\frac{\ddot R}{R}
-\left(2\frac{\dot A}{A}
-\frac{\dot R}{R}\right)\frac{\dot R}{R}\right]+\frac{A}{B}\left[\left(2\frac{A^{\prime}}{A}
+\frac{R^{\prime}}{R}\right)\frac{R^{\prime}}{R}-\left(\frac{B}{R}\right)^2\right],
\label{j2}
\end{eqnarray}
\end{widetext}
where $\stackrel{\Sigma^{(e)}}{=}$ means that both sides of the equation
are evaluated on $\Sigma^{(e)}$.

Comparing (\ref{j2}) with  (\ref{13}) and (\ref{14}) one obtains
\begin{equation}
q\stackrel{\Sigma^{(e)}}{=}P_r.\label{j3}
\end{equation}
Thus   the matching of
(\ref{1})  and (\ref{1int}) on $\Sigma^{(e)}$ implies (\ref{junction1}) and  (\ref{j3}).

As we mentioned in the introduction, the expansion--free models  present an internal vacuum cavity whose boundary surface is denoted by $\Sigma^{(i)}$, then the matching of the Minkowski spacetime within the cavity to the fluid distribution, implies
\begin{equation}
m(t,r)\stackrel{\Sigma^{(i)}}{=}0, \label{junction1i}
\end{equation}
\begin{equation}
q\stackrel{\Sigma^{(i)}}{=}P_r=0.\label{j3i}
\end{equation}

If any of the above matching conditions is not satisfied, then we have to assume a thin shell on the corresponding boundary surface.

\subsection{The Weyl tensor and the complexity factor}
 Some of the solutions exhibited in the next section are obtained from the condition of vanishing complexity factor.  This is  a scalar function intended to measure the degree of complexity of a given fluid distribution \cite{ps1, ps2}, and is related to the so called structure scalars \cite{sc}.

In  the spherically symmetric case the magnetic part of the Weyl tensor  ($C^{\rho}_{\alpha
\beta
\mu}$) vanishes, accordingly the Weyl tensor is  defined by its ``electric'' part
 $E_{\gamma \nu }$ alone, defined by
   \begin{equation}
E_{\alpha \beta} = C_{\alpha \mu \beta \nu} V^\mu V^\nu,
\label{elec}
\end{equation}
whose non trivial components are
\begin{eqnarray}
E_{11}&=&\frac{2}{3}B^2 {\cal E},\nonumber \\
E_{22}&=&-\frac{1}{3} R^2 {\cal E}, \nonumber \\
E_{33}&=& E_{22} \sin^2{\theta},
\label{ecomp}
\end{eqnarray}
where
\begin{widetext}
\begin{eqnarray}
{\cal E}= \frac{1}{2 A^2}\left[\frac{\ddot R}{R} - \frac{\ddot B}{B} - \left(\frac{\dot R}{R} - \frac{\dot B}{B}\right)\left(\frac{\dot A}{A} + \frac{\dot R}{R}\right)\right]+ \frac{1}{2 B^2} \left[\frac{A^{\prime\prime}}{A} - \frac{R^{\prime\prime}}{R} + \left(\frac{B^{\prime}}{B} + \frac{R^{\prime}}{R}\right)\left(\frac{R^{\prime}}{R}-\frac{A^{\prime}}{A}\right)\right] - \frac{1}{2 R^2}.
\label{E}
\end{eqnarray}
\end{widetext}
 Observe that  the electric part of the
Weyl tensor  may be written as
\begin{equation}
E_{\alpha \beta}={\cal E} \left(K_\alpha K_\beta-\frac{1}{3}h_{\alpha \beta}\right).
\label{52}
\end{equation}
As shown in \cite{ps1, ps2} the complexity factor is identified with the scalar function $Y_{TF}$ which defines the trace--free part of the electric Riemann tensor (see \cite{sc} for details).

Thus,
let us define tensor $Y_{\alpha \beta}$ by
\begin{equation}
Y_{\alpha \beta}=R_{\alpha \gamma \beta \delta}V^\gamma V^\delta,
\label{electric}
\end{equation}
which may be expressed in terms of  two scalar functions $Y_T, Y_{TF}$, as
\begin{eqnarray}
Y_{\alpha\beta}=\frac{1}{3}Y_T h_{\alpha
\beta}+Y_{TF}\left(K_{\alpha} K_{\beta}-\frac{1}{3}h_{\alpha
\beta}\right).\label{electric'}
\end{eqnarray}

Then after lengthy but simple calculations, using field equations, we obtain (see \cite{1B} for details)
\begin{eqnarray}
Y_T=4\pi(\mu+3 P_r-2\Pi) , \qquad
Y_{TF}={\cal E}-4\pi \Pi. \label{EY}
\end{eqnarray}

Next, using  (\ref{12}), (\ref{14}), (\ref{15}) with (\ref{17masa}) and (\ref{E}) we obtain
\begin{equation}
\frac{3m}{R^3}=4\pi \left({\mu}-\Pi \right) - \cal{E},
\label{mE}
\end{equation}
and

\begin{equation}
Y_{TF}= -8\pi\Pi +\frac{4\pi}{R^3}\int^r_0{R^3\left(D_R {\mu}-3{q}\frac{U}{RE}\right)R^\prime d\tilde r}.
\label{Y}
\end{equation}
\\
It is worth noticing that due to a different signature, the sign of $Y_{TF}$ in the above equation differs from the sign of the $Y_{TF}$ used in \cite{ps1} for the static case.

For reasons explained in detail in \cite{ps1},  scalar $Y_{TF}$ is the variable identified with the complexity factor.  As it follows from the equations above it may be expressed through the Weyl tensor and the anisotropy of pressure  or in terms of the anisotropy of pressure, the density inhomogeneity and  the dissipative variables.

In terms of the metric functions the scalar $Y_{TF}$ reads

\begin{widetext}
\begin{eqnarray}
Y_{TF}= \frac{1}{A^2}\left[\frac{\ddot R}{R} - \frac{\ddot B}{B} + \frac{\dot A}{A}\left(\frac{\dot B}{B} - \frac{\dot R}{R}\right)\right]+ \frac{1}{ B^2} \left[\frac{A^{\prime\prime}}{A} -\frac{A^{\prime}}{A}\left(\frac{B^{\prime}}{B}+\frac{R^{\prime}}{R}\right)\right] .
\label{itfm}
\end{eqnarray}
\end{widetext}

\section{The transport equation}
In the dissipative case we shall need a transport equation in order to find  the temperature distribution and its evolution. Assuming a causal dissipative theory (e.g. the Israel-- Stewart theory \cite{19nt, 20nt, 21nt}) the transport equation for the heat flux reads
\begin{eqnarray}
\tau h^{\alpha \beta}V^\gamma q_{\beta;\gamma}&+&q^\alpha=-\kappa h^{\alpha \beta}\left(T_{,\beta}+Ta_\beta\right)\nonumber \\&-&\frac{1}{2}\kappa T^2 \left(\frac{\tau V^\beta}{\kappa T^2}\right)_{;\beta} q^\alpha,
\label{tre}
\end{eqnarray}
where $\kappa$ denotes the thermal conductivity, and $T$ and $\tau$ denote temperature and relaxation time respectively.

In the spherically symmetric case under consideration, the transport equation has only one independent component which may be obtained from (\ref{tre}) by contracting with the unit spacelike vector $K^\alpha$, it reads
\begin{equation}
\tau V^\alpha  q_{,\alpha}+q=-\kappa \left(K^\alpha T_{,\alpha}+T a\right)-\frac{1}{2}\kappa T^2\left(\frac{\tau V^\alpha}{\kappa T^2}\right)_{;\alpha} q.
\label{5}
\end{equation}

Sometimes the last term in (\ref{tre}) may be neglected \cite{t8}, producing the so called truncated  transport equation, which reads
\begin{equation}
\tau V^\alpha  q_{,\alpha}+q=-\kappa \left(K^\alpha T_{,\alpha}+T a\right).
\label{5trun}
\end{equation}
\section{The homologous and quasi--homologous conditions}

As mentioned before, in order to specify some of our models we shall impose the condition of vanishing complexity factor. However, for time dependent systems, it is not enough to define the complexity of the fluid distribution. We need also to elucidate  what is the simplest pattern of evolution of the system.

In \cite{ps2} the concept of  homologous evolution was introduced, in analogy with the same concept in classical astrophysics, as to represent the simplest mode of evolution of the fluid distribution (see \cite{hsh} for a discussion on the homologous condition in non--comoving coordinates).

Thus,    the field equation (\ref{13}) written as
\begin{equation}
D_R\left(\frac{U}{R}\right)=\frac{4 \pi}{E} q+\frac{\sigma}{R},
\label{vel24}
\end{equation}
 can be easily integrated to obtain
\begin{equation}
U=\tilde a(t) R+R\int^r_0\left(\frac{4\pi}{E} q+\frac{\sigma}{R}\right)R^{\prime}d\tilde r,
\label{vel25}
\end{equation}
where $\tilde a$ is an integration function, or
\begin{equation}
U=\frac{U_{\Sigma^{(e)}}}{R_{\Sigma^{(e)}}}R-R\int^{r_{\Sigma^{(e)}}}_r\left(\frac{4\pi}{E} q+\frac{\sigma}{ R}\right)R^{\prime}d\tilde r.
\label{vel26}
\end{equation}
If the integral in the above equations vanishes  we have from (\ref{vel25}) or (\ref{vel26}) that
\begin{equation}
 U=\tilde a(t) R.
 \label{ven6}
 \end{equation}

 This relationship  is characteristic of the homologous evolution in Newtonian hydrodynamics \cite{20n, 21n, 22n}. In our case, this may occur if the fluid is shear--free and non dissipative, or if the two terms in the integral cancel each other.

In \cite{ps2}, the term  ``homologous evolution'' was used to characterize  relativistic systems satisfying, besides  (\ref{ven6}), the condition
\begin{equation}
\frac{R_I}{R_{II}}=\mbox{constant},
\label{vena}
\end{equation}
where $R_I$ and $R_{II}$ denote the areal radii of two concentric shells ($I,II$) described by $r=r_I={\rm constant}$, and $r=r_{II}={\rm constant}$, respectively.

The important point that we want to stress here is that (\ref{ven6}) does not imply (\ref{vena}).
Indeed, (\ref{ven6})  implies that for the two shells of fluids $I,II$ we have
\begin{equation}
\frac{U_I}{U_{II}}=\frac{A_{II} \dot R_I}{A_1 \dot R_{II}}=\frac{R_I}{R_{II}},
\label{ven3}
\end{equation}
that implies (\ref{vena}) only if  $A=A(t)$, which by a simple coordinate transformation becomes $A={\rm constant}$. Thus while in the non--relativistic regime (\ref{vena}) always follows from the condition that  the radial velocity is proportional to the radial distance, in the relativistic regime the condition (\ref{ven6}) implies (\ref{vena}), only if the fluid is geodesic.

In \cite{epjc} the homologous condition was relaxed, leading to what was defined as  quasi--homologous evolution,  restricted only by condition  (\ref{ven6}), implying
\begin{equation}
\frac{4\pi}{R^\prime}B  q+\frac{\sigma}{ R}=0.
\label{ch1}
\end{equation}

\section{Shearing expansion--free motion}
If the fluid evolves with vanishing expansion scalar ($\Theta=0$), then from (\ref{5c1}) we have
\begin{equation}
\frac{\dot B}{B}=-2\frac{\dot R}{R}, \label{20}
\end{equation}
or,  integrating
\begin{equation}
B=\frac{g(r)}{R^2}, \label{21}
\end{equation}
where $g(r)$ is an arbitrary function of $r$.

Substituting (\ref{20}) into (\ref{13}) we obtain
\begin{equation}
\frac{{\dot R}^{\prime}}{R}+2\frac{\dot R}{R}\frac{R^{\prime}}{R}-\frac{\dot R}{R}\frac{A^{\prime}}{A}=4\pi qAB, \label{22}
\end{equation}
which can be integrated for ${\dot R}\neq 0$ producing
\begin{equation}
A=\frac{R^2{\dot R}}{\tau_1(t)}\exp\left[-4\pi\int q AB\frac{R}{\dot R}dr\right], \label{23}
\end{equation}
where $\tau_1(t)$ is an arbitrary function of $t$.
With (\ref{21}) and ({\ref{23}), the line element (\ref{1}) becomes
\begin{eqnarray}
ds^2=-\left\{\frac{R^2 {\dot R}}{\alpha}\exp\left[
-4\pi\int qAB\frac{R}{\dot R}dr\right]\right\}^2dt^2 \nonumber\\
+\frac{\alpha^2}{R^4}dr^2+R^2(d\theta^2+\sin^2\theta d\phi^2),
\label{25II}
\end{eqnarray}
which is the general metric for a shearing expansion--free anisotropic dissipative fluid, where  without loss of generality (by reparametrizing $r$ and $t$) we put $g=\tau_1=\alpha$,  where $\alpha$ is a unit constant with dimensions $[r^2]$.

We shall now proceed to find exact analytical  solutions satisfying the expansion--free condition. 

\section{Solutions}

As mentioned before, the expansion--free condition alone is not enough to integrate the field equations. We need to resort to additional  restrictions in order to close the full system of equations. Here, two different families of solutions will be obtained. On the one hand  we shall consider non--geodesic fluids satisfying the vanishing complexity factor condition, complemented with the quasi--homologous condition or  with some simple assumptions on the metric variables.  On the other hand we shall consider geodesic fluids satisfying the vanishing complexity factor condition or  the quasi--homologous condition.

\subsection{Non--geodesic, $Y_{TF}=0$, quasi--homologous evolution  and $\Theta=0$.}
We shall start by assuming  the quasi--homologous condition and the vanishing complexity factor condition for a non--geodesic fluid.
\noindent As mentioned before, imposing  $\Theta=0$, we obtain
\begin{eqnarray}
  \frac{\dot{B}}{B} +\frac{2\dot{R}}{R}&=&0\quad \Rightarrow\quad B=\frac{\alpha}{R^2}\label{ytf1p}.
 \end{eqnarray}

\noindent Let us now impose the quasi--homologous condition (\ref{ven6})
\begin{equation}\label{chomo}
  \dot{R}=\tilde{a}(t)AR.
\end{equation}

Then,  from $\Theta=0$ and $U=\tilde a R$ we have
\begin{eqnarray}
  B &=& \frac{\alpha}{R^2} ,\\
  A &=& \frac{\dot{R}}{\tilde a R}.
\end{eqnarray}

\noindent With the above conditions the  physical variables read
\begin{eqnarray}
8\pi \mu&=&-3{\tilde a}^2-\frac{R^4}{\alpha^2}\left [ \frac{2R^{\prime \prime}}{R}+5\left(\frac{R^\prime}{R}\right)^2 -\frac{\alpha^2}{R^6}   \right],\\
  4\pi q&=&\frac{3{\tilde a}}{\alpha} R R^\prime,\\
  8\pi P_r&=&-3{\tilde a}^2+\frac{R^4}{\alpha^2}\left [\frac{2\dot{R}^{\prime }}{\dot{R}}\frac{R^\prime}{R}-\left(\frac{R^\prime}{R}\right)^2 -\frac{\alpha^2}{R^6} \right],\\
  8\pi P_\bot&=&3{\tilde a}^2+\frac{R^4}{\alpha^2}\left [\frac{\dot{R}^{\prime \prime }}{\dot{R}}+\frac{\dot{R}^\prime}{\dot{R}}\frac{R^\prime}{R}+\left(\frac{R^\prime}{R}\right)^2   \right],
\end{eqnarray}
where we have choosen $\tilde a=constant$.

\noindent Next, condition   $Y_{TF}=0$ produces
\begin{equation}\label{Ytf}
  -3\tilde a^2+\frac{R^4}{\alpha^2}\left [\frac{\dot{R}^{\prime\prime}}{\dot{R}}-\frac{\dot{R}^{\prime}}{\dot{R}} \frac{R^\prime}{R}-\frac{R^{\prime\prime}}{R}+\left(\frac{R^\prime}{R}\right)^2 \right ]=0.
\end{equation}

\noindent In order to find a solution to the above equation, we shall proceed as follows. We shall write $R$ as
\begin{equation}
R=F(\delta_1 r+\delta_2 t+\delta_3)\equiv F(z),\label{fF}
\end{equation}
\noindent where $F$ is an arbitrary function of its argument with dimensions $[r]$, $\delta_1, \delta_2$  are two arbitrary constants with dimensions $[1/r]$, and $\delta_3$ is a dimensionless constant.

Feeding back (\ref{fF}) into  (\ref{Ytf}) we obtain
\begin{equation}\label{ecAFa}
  -3\tilde a^2+\frac{\delta_1 ^2F^4}{\alpha^2}\left[\frac{\frac{\partial^3 F}{\partial z^3}}{\frac{\partial F}{\partial z}}-2\frac{\frac{\partial^2 F}{\partial z^2}}{F}+\left(\frac{\frac{\partial F}{\partial z}}{F}\right)^2 \right]=0.
\end{equation}

\noindent Introducing the variable  $\omega (F)\equiv \frac{\partial F}{\partial z}$ we have

\begin{eqnarray}
\frac{\partial F}{\partial z} &=& \omega, \\
 \frac{\partial ^2 F}{\partial z^2}&=& \omega _F \omega, \\
\frac{\partial ^3 F}{\partial z^3} &=& \omega^2 \omega_{FF}+\omega \omega_F^2,
\end{eqnarray}
\noindent with the help of which  (\ref{ecAFa}) becomes
\begin{equation}\label{ecAF}
  -3\tilde a^2+\frac{\delta_1 ^2F^4}{\alpha^2}\left[\omega \omega_{FF} +\omega_F^2-\frac{2\omega \omega_F}{F}+\frac{\omega^2}{F^2}\right ]=0,
\end{equation}

\noindent where subscript $F$ denotes derivative with respect to $F$, and whose solution reads
\begin{equation}\label{some}
  \omega=\frac{k}{F},\qquad k=\frac{\tilde a\alpha}{\sqrt{2}\delta_1}.
\end{equation}
\noindent Using the above results we may write for $R$
\begin{equation}\label{sR}
  R=\sqrt{2 k(\delta_1 r+ \delta_2 t+\delta_3)},
\end{equation}

\noindent and the physical variables read
\begin{eqnarray}
8\pi \mu&=&-\frac{9}{2}\tilde a^2+\frac{1}{2k(\delta_1 r+\delta_2 t+\delta_3)},\\
  4\pi q&=&\frac{3\tilde a^2}{\sqrt{2}},\\
  8\pi P_r&=&-\frac{9}{2}\tilde a^2-\frac{1}{2k(\delta_1 r+\delta_2 t+\delta_3)},\\
  8\pi P_\bot&=&\frac{9}{2}\tilde a^2.
\end{eqnarray}

\noindent The corresponding expressions for the mass function and  the shear are

\begin{equation}\label{mch}
  m=\frac{\tilde a^2}{4}\left[\sqrt{2k(\delta_1 r+\delta_2 t+\delta_3)}\right]^3+\frac{1}{2}\sqrt{2k(\delta_1 r+\delta_2 t+\delta_3)},
\end{equation}

\begin{equation}\label{sch}
  \sigma=-3\tilde a,
\end{equation}
whereas for the temperature, using (\ref{5trun}) we find 
\begin{equation}\label{Tm3}
  T=\frac{2\tilde a T_0(t)}{\delta_2}\left ( \delta_1 r+\delta_2 t+\delta_3  \right)+\frac{3\tilde a }{8\pi  k},
\end{equation}
where $T_0(t)$ is a function  of integration which in principle may be obtained from the boundary conditions on either boundary surface.

\subsection{Non--geodesic, $Y_{TF}=0$, $\Theta=0$, $A=\gamma B$, $\gamma =constant$.}

The next model will also be obtained by imposing $Y_{TF}=0$, $\Theta=0$, however instead of assuming  the quasi--homologous evolution as in the precedent model, we shall assume  that  $A$ and $B$ are proportional to each other ($A=\gamma B$, with $\gamma =constant$).

\noindent Thus, from the three conditions above we obtain 

\begin{equation}\label{TYtf1}
   \frac{3}{\gamma^2}\frac{\ddot{R}}{R}- \frac{2 R^{\prime \prime}}{R}+\left(\frac{2R^\prime}{R}\right)^2=0.
\end{equation}

\noindent In order to find a solution to  (\ref{TYtf1}), we assume for $R$ the form
\begin{equation}\label{STYtf1}
  R=F(s_1r+s_2t+s_3)\equiv F(z),
\end{equation}
where $s_1, s_2$ are constants with dimensions $[1/r]$ and $s_3$ is a dimensionless constant.

\noindent Replacing  (\ref{STYtf1}) into  (\ref{TYtf1}) we obtain

\begin{equation}\label{ecF}
  \left (  \frac{3s_2^2}{\gamma^2}-2s_1^2 \right)\frac{F_{zz}}{F}+\frac{4s_1^2 F_z^2}{F^2}=0.
\end{equation}

\noindent Next, introducing the variable $y$ defined by

\begin{equation}
y=\frac{F_z}{F}\label{yc},
\end{equation}
\noindent we may write equation (\ref{ecF}) as
\begin{equation}\label{ecFy}
  y_z+\beta_0y^2=0,\qquad with \qquad  \beta_0=\frac{3s_2^2+2s_1^2\gamma^2}{3s_2^2-2s_1^2\gamma^2}.
\end{equation}

\noindent the above equation may be easily integrated, producing
\begin{equation}\label{secFy}
  y=\frac{1}{\beta_0 z+\beta_1},
\end{equation}

\noindent where  $\beta_1$  is a constant of integration. 

Feeding back  (\ref{secFy}) into (\ref{yc}) and integrating once again, we obtain 
\begin{equation}\label{solR}
  F=R=\left ( \frac{\beta_0 z +\beta_1}{\beta_2}  \right )^\frac{1}{\beta_0},
\end{equation}

\noindent where  $\beta_2$ is a new constant of integration with dimensions $[1/r^{\beta_0}]$.

\noindent The physical variables for this model read

\begin{eqnarray}
  8\pi \mu &=& \frac{\left(\frac{\beta_0 z+\beta_1}{\beta_2}\right)^\frac{4}{\beta_0}\left[s_1^2(2\beta_0-7)-\frac{3s_2^2}{\gamma^2}\right]}{\alpha^2(\beta_0 z+\beta_1)^2}\nonumber \\&+&\left(\frac{\beta_2}{\beta_0 z+\beta_1}\right)^\frac{2}{\beta_0},
  \label{m1n1}
  \end{eqnarray}
 \begin{eqnarray}
  4\pi q &=&\frac{\left(\frac{\beta_0 z+\beta_1}{\beta_2}\right)^\frac{4}{\beta_0}s_1s_2(5-\beta_0)}{\gamma \alpha^2(\beta_0 z+\beta_1)^2}, 
 \label{qm1}
 \end{eqnarray}
 \begin{eqnarray}
  8\pi P_r &=&  \frac{\left(\frac{\beta_0 z+\beta_1}{\beta_2}\right)^\frac{4}{\beta_0}\left [\frac{s_2^2(2\beta_0-7)}{\gamma^2}-3s_1^2\right ]}{\alpha^2(\beta_0 z+\beta_1)^2}-\left(\frac{\beta_2}{\beta_0 z+\beta_1}\right)^\frac{2}{\beta_0},\nonumber\\
 \label{pr1m}
 \end{eqnarray}
 \begin{eqnarray}
  8\pi P_\bot &=&  \frac{\left(\frac{\beta_0 z+\beta_1}{\beta_2}\right)^\frac{4}{\beta_0}(1+\beta_0) \left(s_1^2-\frac{s_2^2}{\gamma^2}\right)}{\alpha^2(\beta_0 z+\beta_1)^2},
\label{pt2m}
\end{eqnarray}

\begin{equation}
  m=\frac{\left(\frac{\beta_0 z+\beta_1}{\beta_2}\right)^\frac{7}{\beta_0} \left(\frac{s_2^2}{\gamma^2}-s_1^2\right)}{2\alpha^2(\beta_0 z+\beta_1)^2}+\frac{1}{2}\left(\frac{\beta_0 z+\beta_1}{\beta_2}\right)^\frac{1}{\beta_0},
\end{equation}

\begin{equation}
\sigma=-\frac{3s_2}{\gamma \alpha (\beta_0 z+\beta_1)}\left ( \frac{\beta_0 z+\beta_1}{\beta_2}\right )^\frac{2}{\beta_0},
\end{equation}

\begin{eqnarray}
  T&=&\frac{\left(\frac{\beta_0 z+\beta_1}{\beta_2}\right)^\frac{2}{\beta_0}}{\gamma \alpha}\left[T_0(t)-\frac{\tau s^2_2(5-\beta_0)(2-\beta_0)\left(\frac{\beta_0 z+\beta_1}{\beta_2}\right)^\frac{2}{\beta_0}}{4\pi\kappa \gamma \alpha (1-\beta_0)(\beta_0 z+\beta_1)^2}    \right]\nonumber\\
  &+&\frac{\left(\frac{\beta_0 z+\beta_1}{\beta_2}\right)^\frac{2}{\beta_0}s_2(5-\beta_0)}{4\pi \gamma \kappa \beta_0 \alpha(\beta_0 z+\beta_1)}.
\label{T2m}
\end{eqnarray}

As in the precceeding models, the temperature is obtained  using the truncated transport equation.

\subsection{Non--geodesic, $Y_{TF}=0$, $\Theta=0$, $A=A(r)$, $R=R_1(t)R_2(r)$.}

\noindent We shall now find a solution satisfying conditions  $\Theta=0$ and $Y_{TF}=0$, plus  $A=A(r)$ and the condition that $R$ is a separable function, i.e. $R=R_1(t)R_2(r)$.

From  $\Theta=0$ and $Y_{TF}=0$, we may write

\begin{equation}\label{OTYtf}
  \frac{3}{A^2}\left (\frac{\ddot{R}}{R}-\frac{2\dot{R}^2}{R^2} -\frac{\dot{A}}{A}\frac{\dot{R}}{R} \right )+\frac{1}{B^2}\left ( \frac{A^{\prime \prime}}{A}+\frac{A^\prime}{A}\frac{R^\prime}{R}\right)=0.
\end{equation}

\noindent Imposing further conditions  $A=A(r)$ and  $R=R_1(t)R_2(r)$, we see that a simple solution to (\ref{OTYtf}) reads

\begin{eqnarray} \label{metric3}
  R_1(t) &=& \frac{\nu_0}{ t+\nu_1}, \\
  R_2(r) &=& \nu_2 A^{\nu_3-1}, \\
  A &=& \nu_4\left ( \nu_3r+\nu_5  \right )^{\frac{1}{\nu_3}},
\end{eqnarray}
where $\nu_0, \nu_3$ are dimensionless constants, and $\nu_1, \nu_2, \nu_4, \nu_5$ are constants with dimensions $[r], [r^2], [1/(r^{1/\nu_3})], [r]$, respectively.

\noindent The physical variables for this model read

\begin{eqnarray}
  8\pi \mu &=&\frac{(t+\nu_1)^2\left ( \nu_3r+\nu_5  \right )^{\frac{-2\nu_3+2}{\nu_3}}}{\nu_0^2\nu_2^2\nu_4^{2\nu_3-2}}\nonumber\\
  &-&\frac{\nu_0^4 \nu_4^{4\nu_3-4}\nu_2^4(5\nu_3^2-12\nu_3+7)\left ( \nu_3r+\nu_5 \right )^{\frac{2(\nu_3-2)}{\nu_3}}}{\alpha^2 (t+\nu_1)^4}\nonumber \\ &-&\frac{3\left ( \nu_3r+\nu_5  \right )^{\frac{-2}{\nu_3}}}{\nu_4^2(t+\nu_1)^2},
 \label{mum2}
 \end{eqnarray}
  \begin{eqnarray}
  4\pi q &=&\frac{\nu_2^2\nu_0^2(4-3\nu_3)\nu_4^{2\nu_3-3}\left ( \nu_3r+\nu_5  \right )^{\frac{\nu_3-3}{\nu_3}}}{\alpha (t+\nu_1)^3} ,
\label{qmm2}
\end{eqnarray}
\begin{eqnarray}
  8\pi P_r &=& -\frac{5\left ( \nu_3 r+\nu_5  \right )^{\frac{-2}{\nu_3}}}{\nu_4^2(t+\nu_1)^2}-\frac{(t+\nu_1)^2\left ( \nu_3 r+\nu_5  \right )^{\frac{-2\nu_3+2}{\nu_3}}}{\nu_0^2\nu_2^2\nu_4^{2\nu_3-2}}\nonumber\\
  &+&\frac{\nu_0^4 \nu_4^{4\nu_3-4}\nu_2^4(\nu_3^2-1)\left ( \nu_3 r+\nu_5  \right )^{\frac{2(\nu_3-2)}{\nu_3}}}{\alpha^2 (t+\nu_1)^4 },
  \label{prm2}
  \end{eqnarray}
 \begin{eqnarray}
  8\pi P_\bot &=&  \frac{\nu_0^4 \nu_4^{4\nu_3-4}\nu_2^4(2\nu_3^2-3\nu_3+1)\left ( \nu_3 r+\nu_5  \right )^{\frac{2(\nu_3-2)}{\nu_3}}}{\alpha^2 (t+\nu_1)^4} \nonumber \\&-&\frac{2\left ( \nu_3r+\nu_5 \right )^{\frac{-2}{\nu_3}}}{\nu_4^2(t+\nu_1)^2},
\label{ptm2}
\end{eqnarray}

\begin{equation}\label{shearm}
  \sigma=\frac{3}{\nu_4(t+\nu_1)(\nu_3 r+\nu_5)^\frac{1}{\nu_3}},
\end{equation}

\begin{eqnarray}
  m &=& \frac{\nu_0^3\nu_2^3\nu_4^{3\nu_3-5}(\nu_3 r+\nu_5)^\frac{3\nu_3-5}{\nu_3}}{2(t+\nu_1)^5} +\frac{\nu_0\nu_2\nu_4^{\nu_3-1}(\nu_3 r+\nu_5)^\frac{\nu_3-1}{\nu_3}}{2(t+\nu_1)}\nonumber\\
   &-& \frac{\nu_0^7\nu_2^7\nu_4^{7(\nu_3-1)}(\nu_3-1)^2(\nu_3 r+\nu_5)^\frac{5\nu_3-7}{\nu_3}}{2\alpha^2(t+\nu_1)^7},
\end{eqnarray}

\begin{eqnarray}\label{Tem}
  T&=&\frac{(\nu_3 r+\nu_5)^{\frac{-1}{\nu_3}}}{\nu_4}\left [T_0(t)-\frac{3\tau(4-3\nu_3)}{4\pi\kappa\nu_4(t+\nu_1)^2(\nu_3 r+\nu_5)^\frac{1}{\nu_3}}   \right.\nonumber \\
 &-&\left.\frac{(4-3\nu_3)}{4\pi \kappa\nu_3(t+\nu_1)}\ln (\nu_3 r+\nu_5) \right].
\end{eqnarray}

\subsection{Geodesic models}
 We shall now consider geodesic fluids, for which we have 
\begin{equation}
  A(t,r) =1.
  \label{g1}
  \end{equation}

 Besides, from the expansion--free condition we have

 \begin{equation}
  B(t,r) = \frac{\alpha}{R^2}.
\label{g2}
\end{equation}

From  the above it follows that the general expressions for the physical variables read in this case

\begin{equation}
  8\pi \mu= -\frac{3\dot{R}^2}{R^2}-\frac{R^4}{\alpha^2}\left [2\frac{R^{\prime\prime}}{R}+5\left(\frac{R^\prime}{R}\right)^2 \right ]+\frac{1}{R^2}, 
 \label{gmu1}
 \end{equation}
 \begin{equation}
  4\pi q = \frac{R\dot{R}^\prime}{\alpha}+\frac{2R^\prime \dot{R}}{\alpha}, 
 \label{gqm1}
 \end{equation}
 \begin{equation}
  8\pi P_r =-\frac{2\ddot{R}}{R}-\frac{\dot{R}^2}{R^2}+\left(\frac{R R^\prime}{\alpha}\right)^2-\frac{1}{R^2}, 
  \label{gprm1}
  \end{equation}
  \begin{equation}
  8\pi P_\bot = \frac{\ddot{R}}{R}-\frac{4\dot{R}^2}{R^2}+\frac{R^4}{\alpha^2}\left[ \frac{R^{\prime\prime}}{R}+2\left(\frac{R^\prime}{R}\right)^2 \right ],
\label{gptm1}
\end{equation}

\noindent producing
\begin{equation}\label{tg}
 2 \pi (\mu+P_r+2 P_\bot)=-3\left(\frac{\dot{R}}{R}\right)^2.
\end{equation}

The first model will be obtained from the vanishing complexity factor condition.
\noindent
 Thus, from the condition  $Y_{TF}=0$ we obtain
\begin{equation}\label{g3}
  \frac{\ddot{R}}{R}-\frac{2\dot{R}^2}{R^2}=0\quad \Rightarrow\quad R=\frac{1}{b_1(r)t+b_2(r)}\equiv \frac{1}{b_1(r)\left[t+\frac{b_2(r)}{b_1(r)}\right]},
\end{equation}
where $b_1$ and $b_2$ are two arbitrary functions of their argument, with dimensions $[1/r^2]$ and $[1/r]$ respectively.

Feeding back (\ref{g3}) into (\ref{g2}) we see that by reparametrizying $r$, we may choose without loos of generality $b_1=1/\alpha$. Thus our metric variables become

\begin{equation}
R=\frac{\alpha}{\left[t+\alpha b_2(r)\right]}, \qquad B= \frac{\left[t+\alpha b_2(r)\right]^2}{\alpha}.
\label{g4}
\end{equation}

For this metric,  the physical variables, the mass function and the shear read
\begin{eqnarray}
8\pi \mu&=&-\frac{3}{\left(t+\alpha b_2 \right)^2}+\frac{2\alpha^3 b_2^{\prime \prime}}{\left(t+\alpha b_2 \right)^5}-\frac{9 \alpha^4 (b_2^\prime)^2}{\left(t+\alpha b_2 \right)^6}\nonumber \\&+&\frac{\left(t+\alpha b_2 \right)^2}{\alpha^2},
\label{g5}
\end{eqnarray}
 \begin{eqnarray}
 4\pi q=\frac{4\alpha^2 b_2^\prime}{\left(t+\alpha b_2 \right)^4},
 \label{g6}
 \end{eqnarray}
 \begin{equation}
  8\pi P_r=-\frac{5}{\left(t+\alpha b_2 \right)^2}+\frac{\alpha^4 (b_2^\prime)^2}{\left(t+\alpha b_2 \right)^6}-\frac{\left(t+\alpha b_2 \right)^2}{\alpha^2},
 \label{g7}
 \end{equation}
 \begin{equation}
  8\pi P_\bot=-\frac{2}{\left(t+\alpha b_2 \right)^2}+\frac{4\alpha^4 (b_2^\prime)^2}{\left(t+\alpha b_2 \right)^6}-\frac{\alpha^3 b_2^{\prime \prime}}{\left(t+\alpha b_2 \right)^5},
\label{g8}
\end{equation}

\begin{equation}\label{mgeo}
  m= \frac{\alpha^3}{2(t+\alpha b_2)^5}-\frac{\alpha^7 (b_2^\prime)^2}{2(t+\alpha b_2)^9}+\frac{\alpha}{2(t+\alpha b_2)},
\end{equation}

\begin{equation}\label{sgeo}
  \sigma=\frac{3}{t+\alpha b_2},
\end{equation}

\begin{equation}\label{Tgeo}
  T=T_0(t)+\frac{1}{\pi \kappa(t+\alpha b_2)}\left[1-\frac{2 \tau}{(t+\alpha b_2)}\right].
\end{equation}

A second geodesic model will be obtained from the quasi--homologous condition (\ref{ven6}) which as discussed before, in the geodesic case implies (\ref{vena}), implying in turn that $R$ is a separable function. For this case we obtain
\begin{equation}
R=\frac{g(r)}{t},\qquad B=\frac{t^2}{\alpha},
\label{2gc1}
\end{equation}
where $g(r)$ is an arbitrary function of $r$ with dimensions $[r^2]$.

The physical variables for this model read

\begin{eqnarray}
8\pi \mu=-\frac{3}{t^2}-\frac{\alpha^2}{t^4}\left[\frac{2g^{\prime \prime}}{g}+\left(\frac{g^\prime}{g}\right)^2  \right]+\frac{t^2}{g^2},
\label{og5}
\end{eqnarray}
 \begin{eqnarray}
 4\pi q=-\frac{3\alpha g^\prime}{t^3 g},
 \label{og6}
 \end{eqnarray}
 \begin{eqnarray}
  8\pi P_r=-\frac{5}{t^2}+\frac{\alpha^2}{t^4}\left(\frac{g^\prime}{g}\right)^2-\frac{t^2}{g^2},
 \label{og7}
 \end{eqnarray}
 \begin{eqnarray}
  8\pi P_\bot=-\frac{2}{t^2}+\frac{\alpha^2}{t^4}\frac{g^{\prime\prime}}{g},
\label{og8}
\end{eqnarray}

\begin{equation}\label{mogeo}
  m=\frac{g}{2t}\left[\frac{g^2}{t^4} -\left(\frac{\alpha g^\prime}{t^3}\right)^2+1\right],
\end{equation}

\begin{equation}\label{sogeo}
  \sigma=\frac{3}{t},
\end{equation}

\begin{equation}\label{Togeo}
  T=T_0(t)+\frac{3}{4\pi \kappa t}\left (1-\frac{3\tau}{t}\right )\ln g.
\end{equation}
\\

\section{Conclusions}
The main lesson  we can extract from this work is that  expansion--free condition allows for the obtention  of a wide range of models for  the evolution of spherically symmetric  self-gravitating systems, including   dissipative fluids with anisotropic pressure.

As mentioned in the Introduction, one of the most interesting features of the expansion--free models is the appearance of a vacuum cavity  within the fluid distribution. Whether or not  such models  may be used to describe the formation of voids observed at cosmological scales (see \cite{voids, lid} and references therein), is still an open question. We skip over this issue in a hope of a resolution {\it a posteriori}.

Let us now analyze in some detail the obtained solutions.

The first model satisfies the quasi--homologous and the vanishing complexity factor conditions, and is described by  equations (\ref{sR})--(\ref{Tm3}). Choosing $k, \delta_1, \delta_2, \delta_3>0$ we ensure the positiveness of the  expression within the square root in  (\ref{sR}), which implies because of (\ref{some}) that $\tilde a>0$, and therefore all fluid elements are moving outward. In the limit $t\rightarrow\infty$, the areal radii of all fluid elements  tend to infinity and the fluid distribution becomes a shell. Also, with the above choice we ensure that $R^\prime>0$ avoinding thereby the appearance of shell crossing singularities. In this same limit $8\pi \mu=8\pi P_r=-8\pi P_\bot=-\frac{9 \tilde a^2}{2}$, producing that the inertial mass density $(\mu+P_r)$ is negative. It is worth noticing that the inertial mass density is always negative, not only in the limit  $t\rightarrow\infty$. On the other hand, we see that the expression within the   square bracket  in the ``gravitational term'' in (\ref{3m}) (the first term on the right of (\ref{3m})) is negative as $t\rightarrow\infty$ producing a positive $D_TU$, i.e. such a term  acts as a repulsive force. For sufficiently small values of $\tilde a$ the other parameters of the solution may be chosen such that for some finite time interval the energy density is positive.
Since the heat flux is constant, no contributions from the transient period (terms proportional to $\tau$) appear in the expression of the temperature. Thus this solution might be used to model expansion--free  evolution only for a limited time interval.

The second model also satisfies the vanishing complexity factor condition, but instead of the quasi--homologous evolution we assumed  that the metric functions $A$ and $B$ are proportional.  Its evolution is described by equations (\ref{solR})--(\ref{T2m}). Choosing all the parameters of the solution positive (avoiding thereby shell crossing singularities), then in the limit  $t\rightarrow\infty$, the areal radii of all fluid elements  tend to infinity and the fluid distribution becomes a shell. However in this case depending on the specific values of the parameter $\beta_0$ the behavior of the model may be very different. Indeed, as it follows from (\ref{m1n1})--(\ref{pt2m}) in the limit  $t\rightarrow\infty$, the physical variables tend to zero if $\beta_0>2$ and diverge to infinity if $\beta_0<2$.
 If $\beta_0=2$ the model has a static limit  described by the equation of state $8\pi \mu=8\pi P_r= -\frac{3}{\beta_2^2\alpha^2}\left(s_1^2+\frac{s^2}{\gamma^2}\right)$, with negative energy density and radial pressure. In this case ($\beta_0=2$), both $q$ and $P_\bot$ are constant at all times, implying that the transient effects in temperature vanish, as it is apparent from (\ref{T2m}).
 
 The third solution satisfies  the vanishing complexity factor condition, the metric function $A$ only depends on $r$ and $R$ is a separable function. The full description of this model is provided by equations  (\ref{metric3})--(\ref{Tem}). They describe a collapsing fluid  for which,  in the limit  $t\rightarrow\infty$,  the energy density and the radial pressure diverge and satisfy the equation of state $\mu=-P_r>0$, whereas the heat flux vector and the tangential pressure vanish. In this limit  the transient effects vanish too. An appropriate choice of the parameters allows to obtain well behaved physical variables at least for a finite time interval.
 
 All the three solutions described above are non--geodesic. The next two models instead have a vanishing four--acceleration. 
 
 The first one satisfies the vanishing complexity factor condition, and is described by  equations  (\ref{g3})--(\ref{Tgeo}). This model depicts a collapsing fluid for which as $t\rightarrow\infty$,   the energy density and the radial pressure diverge and satisfy the equation of state $\mu=-P_r$, whereas the heat flux vector and the tangential pressure vanish, and the temperature tends to $T_0$. For sufficiently large (but finite) values of $t$, the energy density is positive the radial pressure is negative, and the fluid evolves almost adiabatically.
 
 Finally, the second geodesic  model is described by equations (\ref{2gc1})--(\ref{Togeo}). In this model the vanishing complexity factor condition is replaced by  the quasi--homologous condition. As the previous model, this one  depicts a collapsing fluid for which as $t\rightarrow\infty$,   the energy density and the radial pressure diverge and satisfy the equation of state $\mu=-P_r$,  the heat flux vector and the tangential pressure vanish, and the temperature tends to $T_0$. Also, for sufficiently large (but finite) values of $t$, the energy density is positive the radial pressure is negative, and the fluid evolves almost adiabatically.
 
 To summarize: the five models presented here might describe some physical realistic situations for  finite time intervals. We notice that neither  of them satisfy the Darmois conditions on either boundary surface, implying that  these are thin shells.
 
We would like to conclude with the following remarks. 
\begin{itemize}
\item The analytical models here presented have the main advantage of simplicity, which allows to use them as toy models for describing the evolution of voids. However they were obtained under specific restrictions, some of which are of purely heuristic nature. In order to get closer to a physically meaningful scenario, one should use some observational data as an input in the solution of the field equations. At this point the best candidate for that purpose appears to be the luminosity profile produced by the dissipative processes within the fluid. Afterwards, it seems unavoidable to  resort to numerical approach in order to solve the field equations .
\item In the first two models the vanishing complexity factor condition leads to two differential equations (\ref{Ytf}), (\ref{TYtf1}) which have been solved analytically resorting  to the heuristic ansatz (\ref{fF}), (\ref{STYtf1}) respectively. Of course, a much more satisfactory way of  doing, would be to solve those equations by using numerical methods. However this would be out of the scope of this work.

\end{itemize}

\begin{acknowledgments}
This  work  was  partially supported by the Spanish  Ministerio de Ciencia, Innovaci\'on, under Research Project No. PID2021-122938NB-I00. 
\end{acknowledgments}

\appendix
\section{Einstein equations}
 Einstein's field equations for the interior spacetime (\ref{1}) are given by
\begin{equation}
G_{\alpha\beta}=8\pi T_{\alpha\beta},
\label{2}
\end{equation}
and its non zero components
read
\begin{widetext}
\begin{eqnarray}
8\pi T_{00}=8\pi  \mu A^2
=\left(2\frac{\dot{B}}{B}+\frac{\dot{R}}{R}\right)\frac{\dot{R}}{R}
-\left(\frac{A}{B}\right)^2\left[2\frac{R^{\prime\prime}}{R}+\left(\frac{R^{\prime}}{R}\right)^2
-2\frac{B^{\prime}}{B}\frac{R^{\prime}}{R}-\left(\frac{B}{R}\right)^2\right],
\label{12} \\
8\pi T_{01}=-8\pi qAB
=-2\left(\frac{{\dot R}^{\prime}}{R}
-\frac{\dot B}{B}\frac{R^{\prime}}{R}-\frac{\dot
R}{R}\frac{A^{\prime}}{A}\right),
\label{13} \\
8\pi T_{11}=8\pi P_r B^2
=-\left(\frac{B}{A}\right)^2\left[2\frac{\ddot{R}}{R}-\left(2\frac{\dot A}{A}-\frac{\dot{R}}{R}\right)
\frac{\dot R}{R}\right]
+\left(2\frac{A^{\prime}}{A}+\frac{R^{\prime}}{R}\right)\frac{R^{\prime}}{R}-\left(\frac{B}{R}\right)^2,
\label{14} \\
8\pi T_{22}=\frac{8\pi}{\sin^2\theta}T_{33}=8\pi P_{\perp}R^2
=-\left(\frac{R}{A}\right)^2\left[\frac{\ddot{B}}{B}+\frac{\ddot{R}}{R}
-\frac{\dot{A}}{A}\left(\frac{\dot{B}}{B}+\frac{\dot{R}}{R}\right)
+\frac{\dot{B}}{B}\frac{\dot{R}}{R}\right]\nonumber \\
+\left(\frac{R}{B}\right)^2\left[\frac{A^{\prime\prime}}{A}
+\frac{R^{\prime\prime}}{R}-\frac{A^{\prime}}{A}\frac{B^{\prime}}{B}
+\left(\frac{A^{\prime}}{A}-\frac{B^{\prime}}{B}\right)\frac{R^{\prime}}{R}\right].\label{15}
\end{eqnarray}
\end{widetext}
The component (\ref{13}) can be rewritten with (\ref{5c1}) and
(\ref{5b}) as
\begin{equation}
4\pi qB=\frac{1}{3}(\Theta-\sigma)^{\prime}
-\sigma\frac{R^{\prime}}{R}.\label{17a}
\end{equation}

\section{Dynamical equations}

The non trivial components of the Bianchi identities, $T^{\alpha\beta}_{;\beta}=0$, from (\ref{2}) yield
\begin{widetext}
\begin{eqnarray}
T^{\alpha\beta}_{;\beta}V_{\alpha}=-\frac{1}{A}\left[\dot { \mu}+
\left( \mu+ P_r\right)\frac{\dot B}{B}
+2\left( \mu+P_{\perp}\right)\frac{\dot R}{R}\right]
-\frac{1}{B}\left[ q^{\prime}+2 q\frac{(AR)^{\prime}}{AR}\right]=0, \label{j4}\\
T^{\alpha\beta}_{;\beta}K_{\alpha}=\frac{1}{A}\left[\dot { q}
+2 q\left(\frac{\dot B}{B}+\frac{\dot R}{R}\right)\right]
+\frac{1}{B}\left[ P_r^{\prime}
+\left(\mu+ P_r \right)\frac{A^{\prime}}{A}
+2( P_r-P_{\perp})\frac{R^{\prime}}{R}\right]=0, \label{j5}
\end{eqnarray}
\end{widetext}
this last equation   can be cast into the form

\begin{widetext}
\begin{eqnarray}
\left( \mu+ P_r\right)D_TU
=-\left(\mu+ P_r \right)
\left[\frac{m}{R^2}
+4\pi  P_r R\right]
-E^2\left[D_R  P_r
+2(P_r-P_{\perp})\frac{1}{R}\right]
-E\left[D_T q+2 q\left(2\frac{U}{R}+\sigma\right)\right].
\label{3m}
\end{eqnarray}
\end{widetext}

\end{document}